\documentclass[prl,twocolumn,showpacs,nobibnotes,floatfix,superscriptaddress]{revtex4}

\usepackage{graphicx,eucal,amsfonts}

\begin{document}

\title{Entanglement Generation of Nearly-Random Operators}

\author{Yaakov S. Weinstein}
\thanks{To whom correspondence should be addressed. \\Present address: Quantum Information Science Group, 
MITRE, Eatontown, NJ, 07724}
\email{weinstein@mitre.org}
\author{C. Stephen Hellberg}
\email{hellberg@dave.nrl.navy.mil}
\affiliation{Center for Computational Materials Science, Naval Research Laboratory, Washington, DC 20375}

\begin{abstract}
We study the entanglement generation of operators whose statistical properties 
approach those of random matrices but are restricted in some way. These 
include interpolating ensemble matrices, where the interval of the 
independent random parameters are restricted, pseudo-random operators, 
where there are far fewer random parameters than required for random matrices,
and quantum chaotic evolution. Restricting randomness in different ways allows us 
to probe connections between entanglement and randomness. We comment on which 
properties affect entanglement generation and discuss ways of efficiently producing 
random states on a quantum computer. 
\end{abstract}

\maketitle

Highly entangled, random, quantum states play a central role in many aspects 
of quantum information processing (QIP). Protocols enabled by random quantum 
states include superdense coding \cite{Aram}, remote state preparation 
\cite{Bennet}, data hiding schemes \cite{Hayden}, and single spin 
measurement \cite{Paola}. Random states are produced from computational
basis states by applying random unitary operators. However, the 
implementation of operators randomly drawn from the circular unitary 
ensemble (CUE), the space of all unitaries, is inefficient.

Independent of its usefulness in QIP, entanglement is a uniquely quantum phenomenon. 
Entanglement is a conjectured signature of quantum chaos \cite{L1,FNP,MS,WGSH,TFM} 
and plays an important role in studies of decoherence \cite{Z1} and measurement. 
Understanding entanglement allows us to better exploit it as a QIP resource
and provides insight into the working of quantum mechanics.

In this paper we study the entanglement production of operator classes that 
approach, but do not properly cover, CUE. The purpose of this is two-fold.
First, it allows us to explore the relationship between randomness and 
entanglement and investigate which statistical properties of randomness lead to 
entanglement production. Second, some of the operators explored here may be 
implemented efficiently. Thus, this study is an exercise of how best to produce 
highly entangled random states. 

The first class of operators we explore are ensembles of random matrices 
which interpolate between integrable and CUE \cite{Zyc3}. These operators 
require the same number of random parameters as CUE but the 
parameters are drawn from restricted intervals. The second class is 
pseudo-random (PR) operators \cite{RM,QCARM,CRM}, possibly efficient 
substitutes for random operators in QIP protocols. These operators fall far 
short of the requisite number of random parameters when compared to CUE 
operators. The third class is quantum analogs of classically chaotic systems. 
These operators are the most restricted in terms 
of random parameters but are known to exhibit certain statistical properties 
of random matrices \cite{Haake} and can be efficiently implemented on a 
quantum computer.

As a practical measure of multi-partite entanglement, we use the average bipartite 
entanglement between each qubit and the rest of the system \cite{Meyer,Bren2},
\begin{equation}
Q(|\psi\rangle) = \frac{4}{n}\sum^n_{j = 1}D(|\tilde{u}_j\rangle,|\tilde{v}_j\rangle) = 2-\frac{2}{n}\sum^n_{j=1}Tr[\rho_j^2],
\end{equation}
where $|\psi\rangle = |0\rangle_j\otimes|\tilde{u}_j\rangle+|1\rangle_j\otimes|\tilde{v}_j\rangle$, $D(|\tilde{u}_j\rangle,|\tilde{v}_j\rangle)$ 
is the norm-squared of the wedge product between $|\tilde{u}_j\rangle$ and 
$|\tilde{v}_j\rangle$, and $\rho_j$ is the reduced density matrix of qubit $j$. 
We apply matrices from the above classes to computational basis states
and the average entanglement produced as a function of time (number of iterations), 
$\langle Q(t) \rangle$, is compared to the CUE average entanglement 
$\langle Q \rangle_{CUE}= (N-2)/(N+1)$ \cite{Scott}, where $N$ is the Hilbert 
space dimension. Other entanglement measures, specifically the concurrence 
between the two most significant qubits and linear entropy between the two $N/2$ 
dimensional subspaces, exhibit behavior 
similar to $Q$ for the operators explored here.

The statistical properties we examine are the level or number variance, the 
randomness of the eigenvectors, and the matrix element distribution. The 
number variance measures a two-point eigenvalue 
correlation function which, for many dynamical systems, is known to deviate 
from CUE at long range due to short periodic orbits \cite{AS}. Thus, the 
number variance provides insight into entanglement generation as a 
function of time since periodic orbits will cause deviations from CUE. 
In the limit of large $N$, the CUE number variance is \cite{Mehta}
$\Sigma^2_{CUE}(L) = \frac{1}{\pi^2}\left(\ln(2\pi L) + 1 + \gamma\right)$,
where $\gamma \simeq .577$ is the Euler constant.

A lower bound for the asymptotic bipartite entanglement production with respect to time, 
$S_{asy}$, is the bipartite entanglement of the systems' eigenvectors, $S_{eig}$ minus one \cite{Dem}. 
This result can be extended to $Q$ since it is an average of bipartite entanglements, 
$Q_{asy} \geq 2Q_{eig}-1$. Let $c^l_k$ denote the $k$th component of the 
$l$th system eigenvector. The distribution of amplitudes, $\eta = |c^l_k|^2$, for CUE
eigenvectors in the limit $N \rightarrow \infty$ and after rescaling to unit 
mean is $P_{CUE}(y) = e^{-y}$, where $y = N\eta$ \cite{Zyc}.

When applying an operator to a computational basis state the resulting state 
is a column of the applied operator. Repeated applications are simply powers
of the column. Thus, an operator's matrix elements, the resulting 
state elements, play a central role in the amount of entanglement generated. 
This is seen by writing the average $Q$ over all states in terms of $|c_i|^2$, 
the state elememt amplitudes
\begin{equation}
\langle Q \rangle = 4\big(\sum^{N/2}_{m = 1}\sum^N_{n = \frac{N}{2}+1}\langle|c_m|^2|c_n|^2\rangle - \sum^{N/2}_{q = 1}\langle |c_q|^2|c_{q+\frac{N}{2}}|^2\rangle\big).
\label{Qavg}
\end{equation}
CUE matrices can be generated by multiplying eigenvectors of a 
Gaussian unitary ensemble (GUE) Hermitian matrix by random phases and 
using the resulting vectors as matrix columns \cite{Zyc2}. Since the eigenvector 
distribution of CUE and GUE are the same \cite{Haake} and multiplication by a 
phase does not change the amplitude of the elements, $P_{CUE}(x)$, the 
distribution of the rescaled amplitude of CUE matrix elements $x$, is equal 
to $P_{CUE}(y)$. The closeness of an operator's matrix element amplitude 
distribution to that of CUE indicates of how much entanglement the 
operator can generate. 

The interpolating ensembles are a one-parameter interpolation between 
diagonal matrices with uniform, independently distributed elements, and CUE
\cite{Zyc3}. They are constructed based on the Hurwitz parameterization of 
CUE matrices. The CUE construction starts with elementary unitary 
transformations, $E^{(i,j)}(\phi,\psi,\chi)$, with non-zero elements 
\cite{Zyc2,corr}
\begin{eqnarray}
\label{E1}
E_{kk}^{(i,j)} &=& 1, \;\;\;\; k = 1, ... , N, \;\;\;\; k \neq i,j \nonumber\\
E_{ii}^{(i,j)} &=& e^{i\psi}\cos\phi, \;\;\;\;\;\;\;\;\; E_{ij}^{(i,j)} = e^{i\chi}\sin\phi \nonumber\\
E_{ji}^{(i,j)} &=& -e^{-i\chi}\sin\phi, \;\;\;\; E_{jj}^{(i,j)} = e^{-i\psi}\cos\phi
\end{eqnarray}
which are used to form $N-1$ composite rotations
\begin{eqnarray}
E_1 &=& E^{(N-1,N)}(\phi_{01},\psi_{01},\chi_1) \nonumber\\
E_2 &=& E^{(N-2,N-1)}(\phi_{12},\psi_{12},0)E^{(N-1,N)}(\phi_{02},\psi_{02},\chi_2) \nonumber\\
\dots\nonumber\\
E_{N-1} &=& E^{(1,2)}(\phi_{N-2,N-1},\psi_{N-2,N-1},0)\times\nonumber\\
 & & E^{(2,3)}(\phi_{N-3,N-1},\psi_{N-3,N-1},0)\times\nonumber\\
 &\dots& E^{(N-1,N)}(\phi_{0,N-1},\psi_{0,N-1},\chi_{N-1})
\end{eqnarray}
and, finally, $U_{CUE} = e^{i\alpha}E_1E_2\dots E_{N-1}$. 
Angles $\psi$, $\chi$, and $\alpha$ are drawn uniformly from the intervals
\begin{equation}
\label{E3}
0\leq \psi_{rs} \leq 2\pi, \;\;\;\;\;\; 0\leq \chi_{s} \leq 2\pi, \;\;\;\;\;\;
0\leq \alpha \leq 2\pi,
\end{equation}
and $\phi_{rs} = \sin^{-1}({\xi_{rs}}^{1/(2r+2)})$, with $\xi_{rs}$ 
drawn uniformly from 0 to 1. The $2\times2$ block $E^{(i,j)}_{m,n}$ with 
$m,n = i,j$ and $r = 0$ is a random SU(2) rotation with respect to the Haar 
measure. Interpolating ensemble construction is the same with the angles 
drawn from constricted intervals
\begin{equation}
\label{delta}
0 \leq \psi_{rs} \leq 2\pi\delta, \;\;\;\;\;\; 0\leq \chi_s \leq 2\pi\delta, \;\;\;\;\;\; 0 \leq \alpha \leq 2\pi\delta, 
\end{equation}
with $\phi_{rs} = \sin^{-1}(\delta{\xi_{rs}}^{1/(2r+2)})$ and
$\xi_{rs}$ drawn from 0 to 1. The whole is multiplied by a diagonal matrix of 
random phases drawn uniformly from 0 to $2\pi$. The parameter $\delta$ 
ranges from 0 to 1 and provides a smooth transition of certain statistical 
properties between the diagonal circular Poisson ensemble and CUE 
\cite{Zyc3}. 

\begin{figure}
\includegraphics[width=4.275cm]{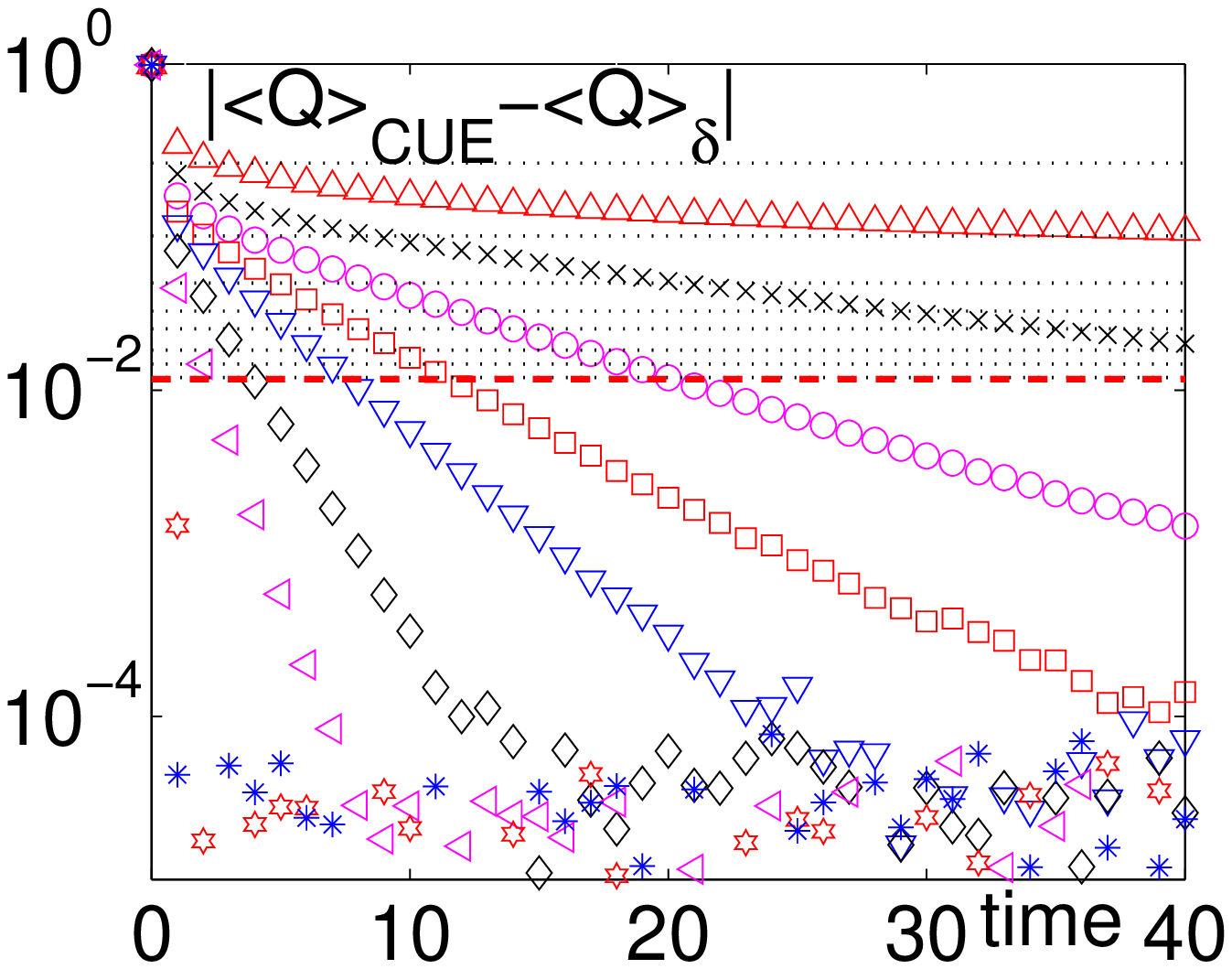}
\includegraphics[width=4.275cm]{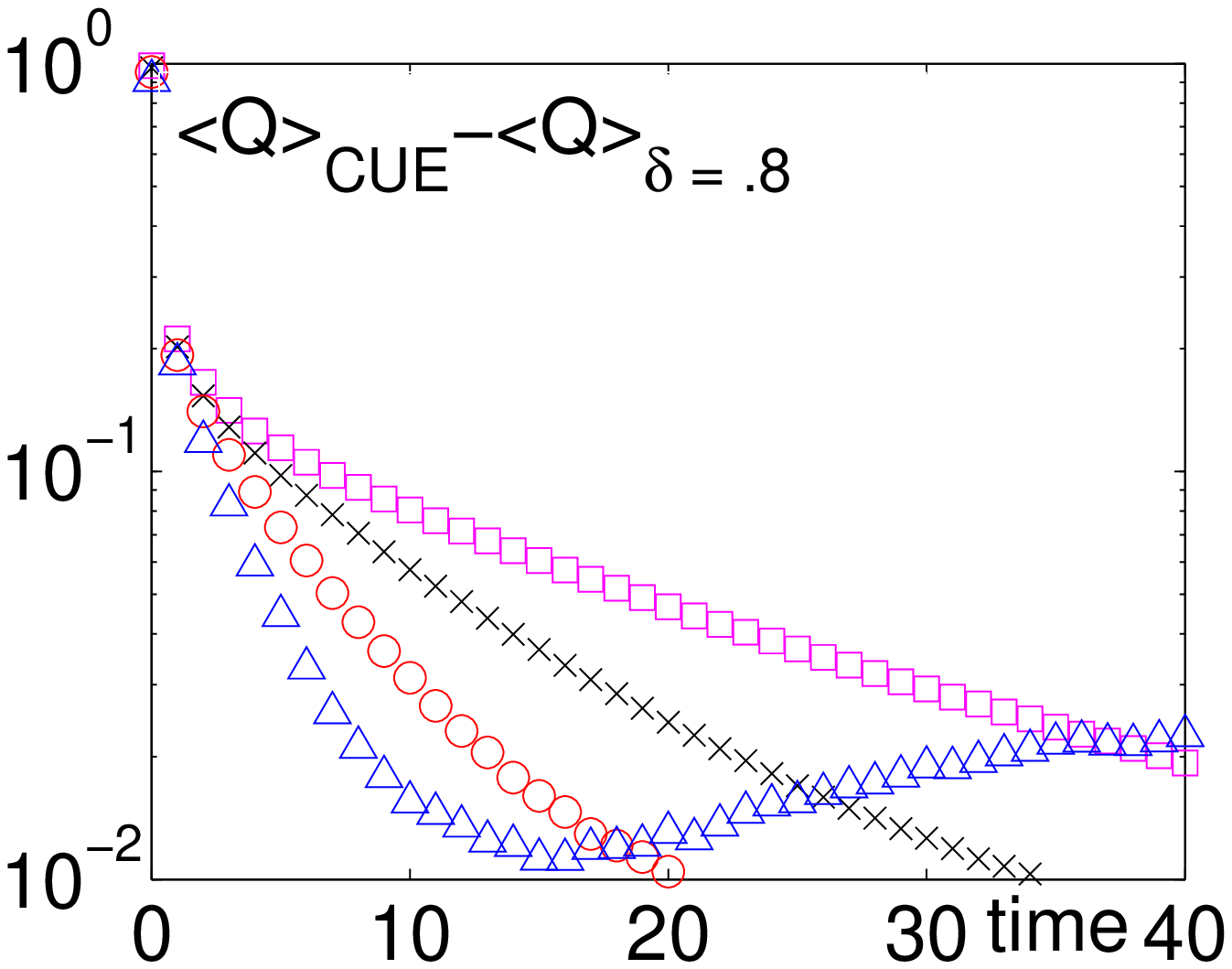}
\caption{\label{CUEd}
(Color online) Absolute value of difference between the CUE average entanglement, 
$\langle Q \rangle_{CUE}$ and the average $Q$ for interpolating ensemble operators, 
$\langle Q \rangle_{\delta}$, applied $t$ times to each computational basis state. 
(left) $N = 256$, $\langle Q\rangle_{CUE} = .9883$ and $\delta = .5$ ($\bigtriangleup$), 
.8 ($\times$), .9 ($\bigcirc$), .94 ($\square$), .96 ($\bigtriangledown$), 
.98 ($\diamond$), .99 (left-triangles), .999 (six-pointed stars), and .9999 ($\ast$). 
The dotted lines show $Q_{asy}$ determined by the eigenvectors of each set of operators. 
The dashed line shows the lower-bound for CUE eigenvectors.  
(right) $N = 256$ ($\square$), 128 ($\times$), 64 ($\bigcirc$), and 32 
($\bigtriangleup$). As $N$ decreases the difference between 
$\langle Q \rangle_{\delta}$ and $\langle Q \rangle_{CUE}$ as a function of 
time goes from power-law to exponential (for $\delta = .8$) and the exponential
rate increases. This is due to the increased randomness of the matrix 
elements.}
\end{figure}

For our purposes the interpolating ensembles have the same number 
of random parameters as CUE matrices, drawn, however, from restricted 
intervals. We stress that $\delta$ restricts all $N^2$ independent variables, 
such that even ensembles of the highest $\delta$ used here cover only an extremely 
small fraction of CUE space. Figure \ref{CUEd} shows the difference between 
$\langle Q \rangle_{\delta}$, and $\langle Q \rangle_{CUE}$ as a function of time 
and $\delta$. For $\delta \gtrsim .96$, $\langle Q \rangle_{\delta}$ 
approaches $\langle Q \rangle_{CUE}$ as an exponential whose rate 
decreases with decreasing $\delta$. For $\delta \lesssim .5$, the approach is a 
power-law. For constant $\delta$, $\langle Q \rangle_{\delta}$ 
approaches its asymptotic value faster for lower $N$ while the difference 
between the asymptotic value and $\langle Q \rangle_{CUE}$ increases for 
lower $N$.

The next class of operators we investigate is PR operators 
\cite{RM,CRM,QCARM}, potentially efficient replacements of CUE operators 
for QIP. To implement a PR operator apply $m$ iterations of the $n$ qubit 
gate: random SU(2) rotation, Eqs. (\ref{E1}) and (\ref{E3}), 
to each qubit, evolve the system via all nearest neighbor couplings \cite{RM}. 
The coupling operator is
$U_{nnc} = e^{i(\pi/4)\sum^{n-1}_{j=1}\sigma_z^j\otimes\sigma_z^{j+1}}$,
where $\sigma_z^j$ is the $j$th qubit $z$-direction Pauli spin operator.
The random rotations are different for each qubit and each iteration. 
After the $m$th iteration, a final set of random rotations is applied. 

The total number of random parameters used to create a PR 
operator is $3n(m+1)+1$ where $n$ is the number of qubits. This is compared 
to $2^{2n} = N^2$ random parameters needed for a CUE matrix. Unless $m$ is 
exponential in $n$ these operators cannot cover CUE simply because there are 
too few random parameters. 

\begin{figure}
\includegraphics[width=4.275cm]{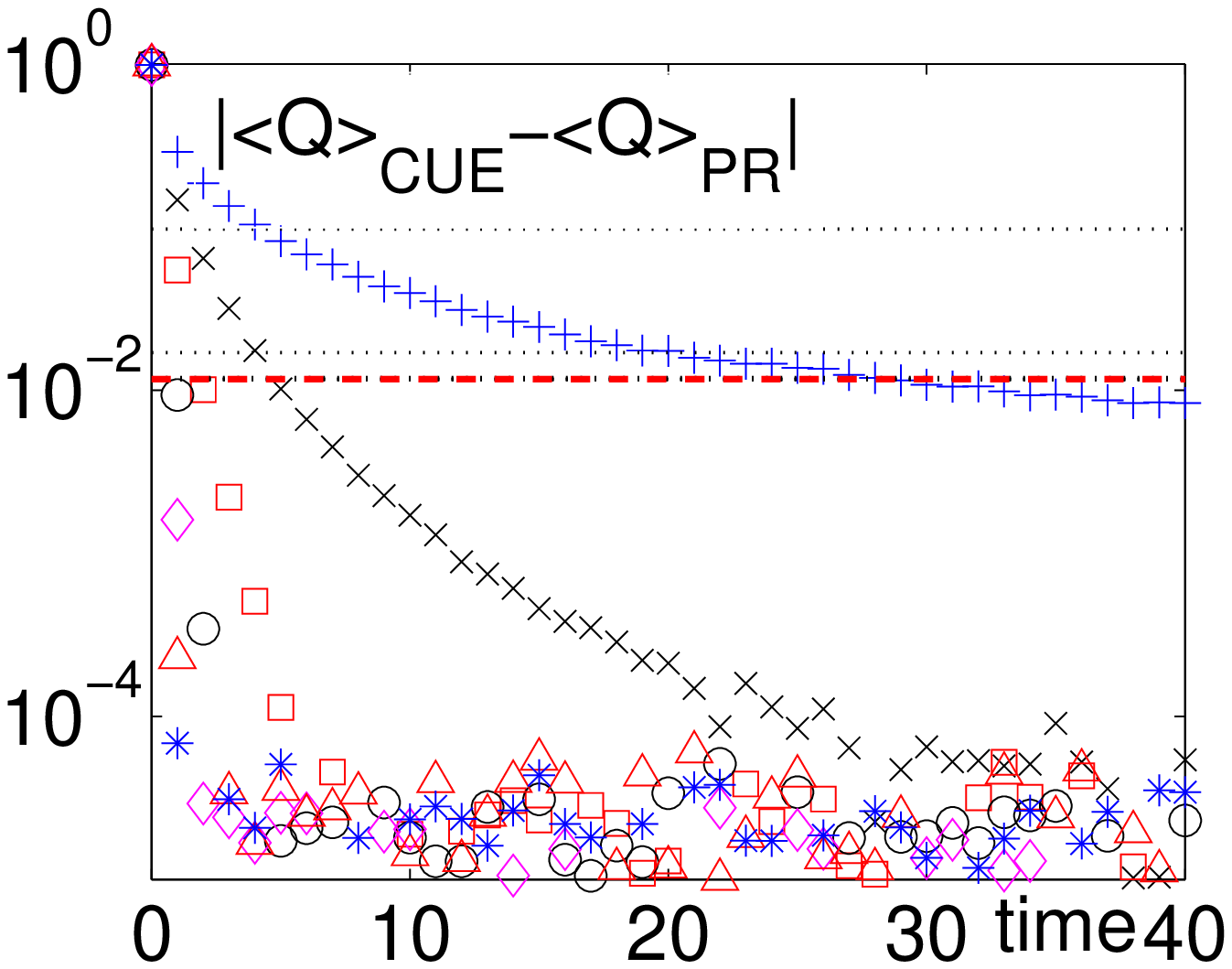}
\includegraphics[width=4.275cm]{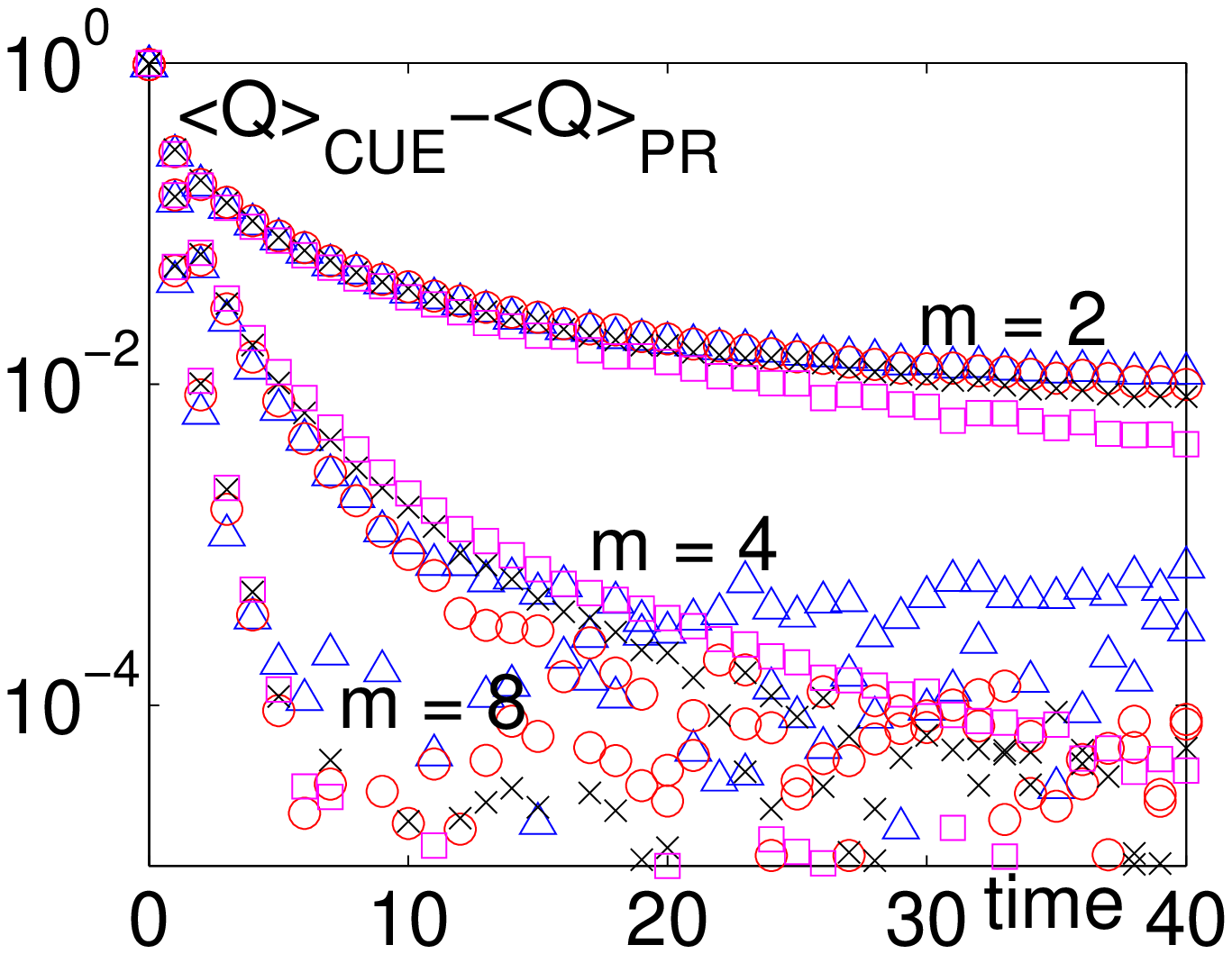}
\caption{\label{PR}
(Color online) Absolute value of difference between $\langle Q\rangle_{CUE}$ 
and $\langle Q(t)\rangle_{m}$. (left) $n = 8$ qubits, $m = 2$ (+), 4 
($\times$), 8 ($\square$), 16 ($\circ$), 24 ($\diamond$), 32 
($\bigtriangleup$), and 40 ($\ast$). Each average is taken 
over 100 operators applied to all $N = 256$ computational basis states. 
The dotted lines show $Q_{asy}$ as determined from the 
eigenvectors. Note that the lower bound for $m \geq 16$ operators is higher
than the lower bound of random eigenvectors. 
(right) The same for $n = 9$ ($\square$), 8 ($\times$), 7 ($\bigcirc$),
and 6 ($\bigtriangleup$). The entanglement production compared to CUE changes 
only slightly as a function of $n$.
}
\end{figure}

The absolute value of the difference between $\langle Q \rangle_{CUE}$ and 
$\langle Q \rangle_{m}$ for $n = 8$ PR operators, 
as a function of time is shown in Fig. \ref{PR}. As with the interpolating 
ensemble operators, $\langle Q \rangle_{m}$ 
approaches $\langle Q \rangle_{CUE}$ as a power-law, for low values of 
$m$, less coverage of CUE, and as an exponential for greater 
$m$. There are a number of interesting features in this plot. First, for 
values of $m$ where $\langle Q(t)\rangle_{m}$ approaches 
$\langle Q\rangle_{CUE}$ exponentially, the average entanglement fluctuates 
around $\langle Q\rangle_{CUE}$ after the exponential saturates (Fig.
\ref{PR} plots absolute value). $\langle Q\rangle_{40}$ converges immediately into 
these fluctuations so increasing $m$ beyond 40 will not increase entanglement 
generation. Also, for operators exhibiting exponential convergence, an operator 
with $m = m_1$ at time $t_1$ has approximately the same $\langle Q\rangle$ as 
an operator with $m_2 = \alpha m_1$ at time $t_2 = t_1/\alpha$. For example, 
$\langle Q(t = 1)\rangle_{m = 24}$ is about equal to  
$\langle Q(t = 3)\rangle_{m = 8}$. This is not the case for 
operators that exhibit non-exponential decay. To create states 
with $\langle Q\rangle \simeq \langle Q\rangle_{CUE}$ one can apply an 
$m = 40$ operator once or an $m = 8$ operator 5 times. These procedures take 
the same amount of time, but the $m = 8$ operator requires fewer random 
parameters. Thus, the number of independent variables needed to create 
entanglement  $\simeq \langle Q\rangle_{CUE}$ in a reasonable time is 
the number required for the lowest $m$ that gives exponential convergence 
(for $n = 8$ this is approximately $m = 8$). Unlike the interpolating ensemble
matrices, the behavior of $Q$ as a function of time barely changes with $N$.

The above shows that one can create states with CUE levels of multi-partite 
entanglement though only a small portion of CUE is covered. An $m = 8$ 
PR operator, for example, has only $193/256^2 = .3\%$ of the 
random parameters needed for CUE operators but can generate CUE levels of 
entanglement by iterating the operator 5-6 times. 

The interpolating ensemble operators and PR operators lead 
to similar average entanglement generation behavior as a function of time. 
For both the entanglement approaches $\langle Q \rangle_{CUE}$ as a power-law 
and, as the operators cover more of CUE, an exponential. A priori, there 
is no reason that different restrictions on CUE should give rise 
to similar average entanglement behavior, especially since the distributions 
of $Q$ after one iteration are very different for the two types of operators
(not shown).

\begin{figure}
\includegraphics[width=4.275cm]{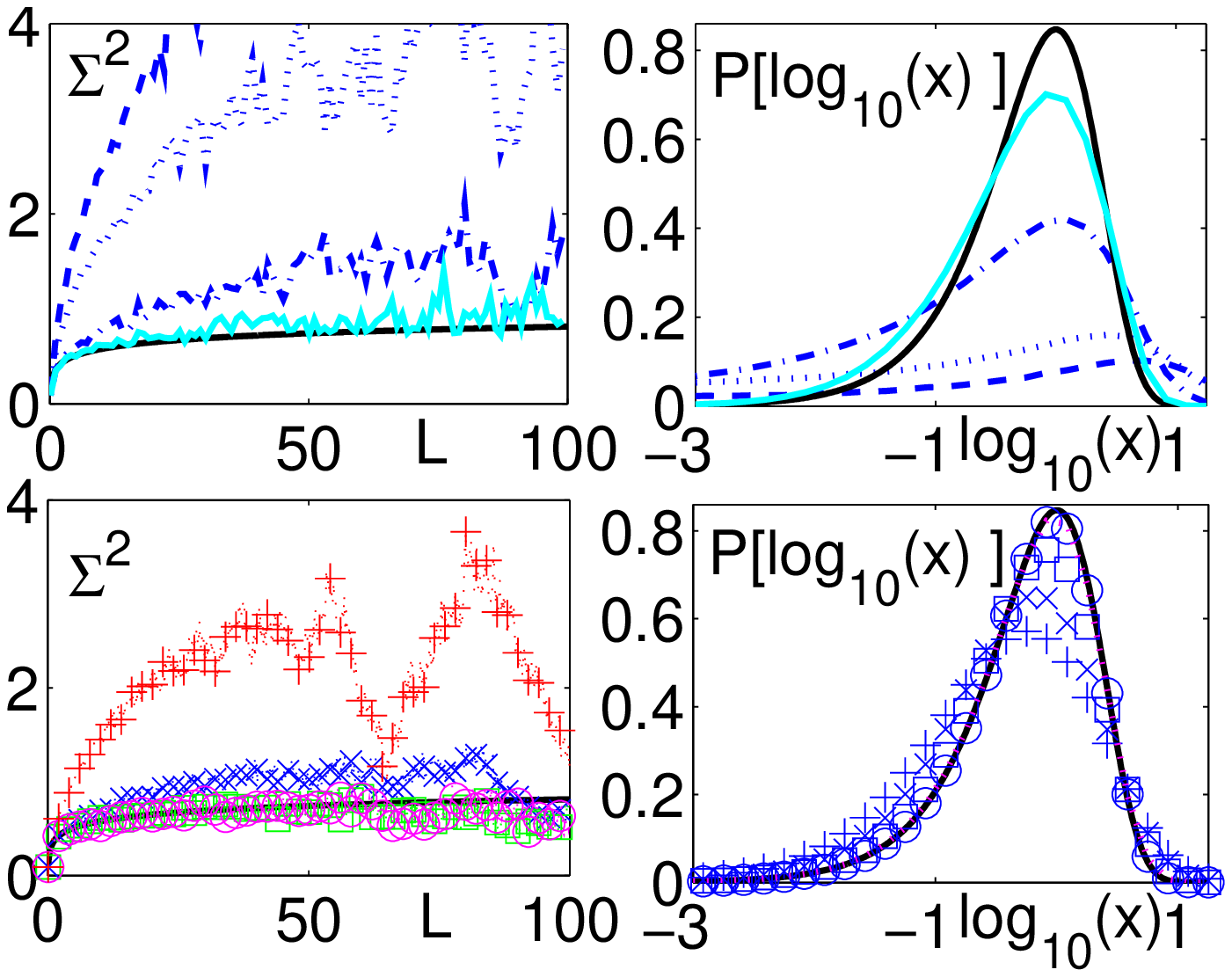}
\includegraphics[width=4.275cm]{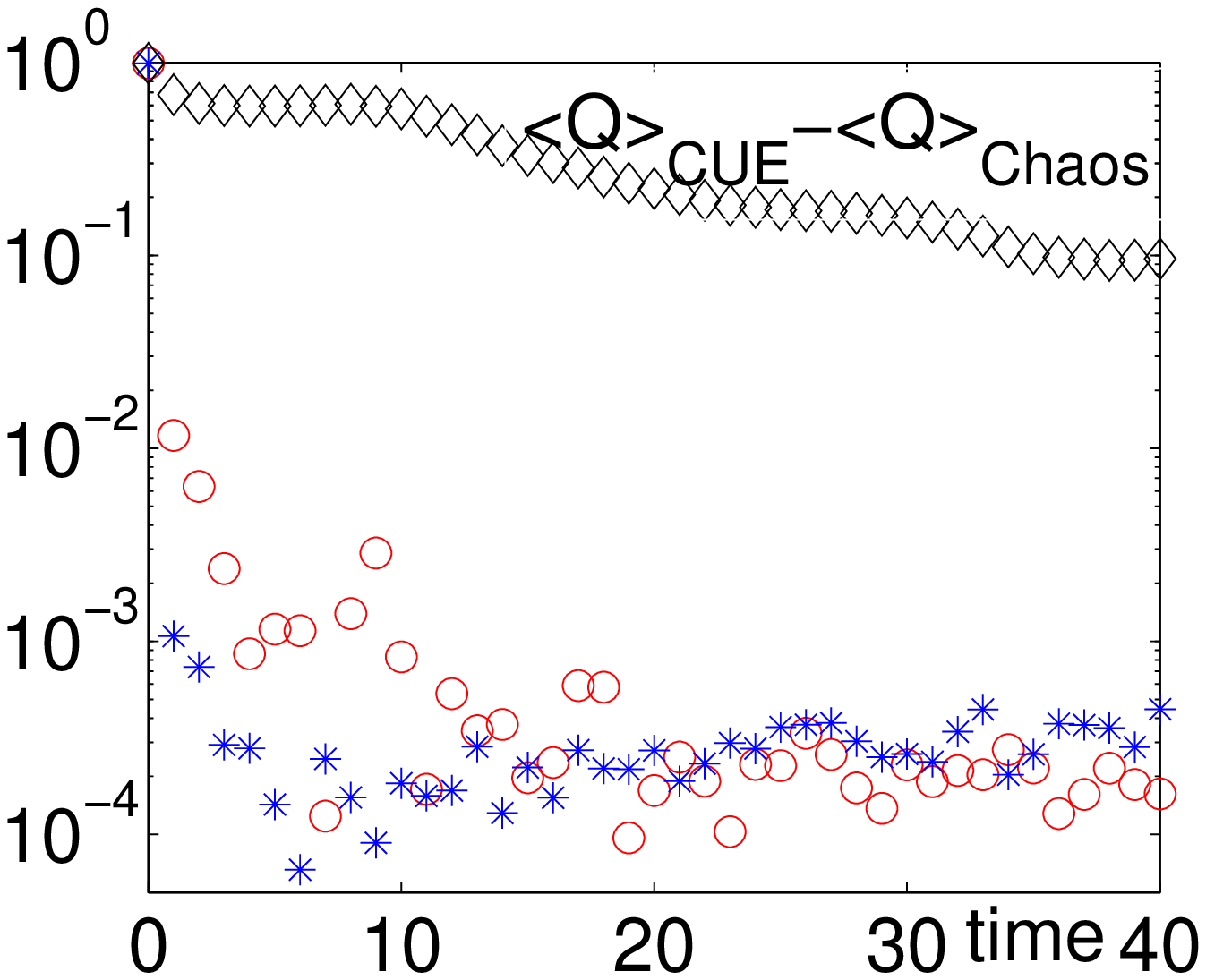}
\caption{\label{comp}
(Color online) (left) Number variance (left) and matrix element distribution 
(right), for interpolating ensemble (top) and PR operators (bottom).
Shown are statistics for interpolating ensembles with $\delta = .1$ (dashed 
line), .5 (dotted line), .9 (chained line) and .98 (light solid line). For 
the PR operators statistics are shown for $m = 2$ (+), 4 ($\times$),
8 ($\square$), and 16 ($\bigcirc$). As $m$ and $\delta$ are increased the 
statistical properties approach CUE (solid line). The rate of approach 
for the matrix element distribution is slower than other statistics. 
(right) Absolute value of difference between $\langle Q\rangle_{CUE}$ and 
$\langle Q(t)\rangle$ of the quantum baker's map ($\diamond$) and sawtooth map 
for non-integer kick strength $0 < k_{saw} < 5$ ($\bigcirc$) and 
Harper's map for $1 \leq \gamma_{H} \leq 6$ ($\ast$).    
}
\end{figure}

In light of these results Fig. \ref{comp} shows how various statistical properties
relate to entanglement production with the aim: to
explain why the entanglement generation approaches $\langle Q \rangle_{CUE}$ 
as a power-law or exponential, and why $\langle Q \rangle_{\delta}$ is 
dependent on $N$ while $\langle Q \rangle_{PR}$ is not. 

Based on our numerical investigations, the number variance determines the 
approach of $\langle Q\rangle$ to $\langle Q\rangle_{CUE}$. For the 
interpolating ensembles with $\delta < .9$ and low $m$ PR operators
the number variance diverges almost immediately from 
$\Sigma^2_{CUE}(L)$. For values of $\delta$ and $m$ where $\langle Q \rangle$ 
approaches $\langle Q\rangle_{CUE}$ exponentially as a function of time, 
$\Sigma^2_{\delta}(L)$ follows CUE faithfully even for large $L$. 
In other words, operators for which $\langle Q\rangle$ 
approaches $\langle Q\rangle_{CUE}$ as a power-law do not follow 
$\Sigma^2_{CUE}(L)$, while those that approach $\langle Q\rangle_{CUE}$ as an
exponential follow $\Sigma^2_{CUE}(L)$ up to long range correlations, 
corresponding to short time.

As expressed in Eq. \ref{Qavg}, the matrix element distribution 
(Fig. \ref{comp}) is key in determining entanglement generation. The matrix 
element distribution approaches the CUE distribution at 
a rate similar to that of the one iteration entanglement distribution, 
$P(Q)$ (not shown). This rate of convergence is slower than that of 
other explored statistics. Second, the matrix element distributions explains
the behavior of the entanglement generation as a function of $N$. For the 
interpolating ensemble operators, the eigenvector and eigenvalue statistics 
are practically unchanged with $N$ \cite{Zyc3}. The matrix element 
distribution, however, is strongly dependent on $N$ as is the entanglement
generation. For $\delta = .9$ operators, the matrix element distribution is 
fully random for $N = 8$ but deviates as $N$ increases. For the PR 
operators the eigenvector and eigenvalue statistics depend only slightly on 
$N$, as with the matrix element distribution. The entanglement generation 
is also essentially unchanged.


The final class of operators we explore are quantum chaotic operators. While a 
full analysis is beyond the scope of this paper we discuss them in 
comparison to the other operators. Figure \ref{comp} shows entanglement 
generation as a function of time for the quantum baker's map \cite{BV} and an 
ensemble of chaotic sawtooth \cite{saw}, and Harper maps \cite{Harper}. The 
convergence to $\langle Q \rangle_{CUE}$ is between exponential and Gaussian but 
the asymptotic value is lower than the other operators. The chaotic maps require 
on order $n^2$ \cite{BV} or $n^3$ \cite{saw} gates. The 
PR operators require $n - 1$ coupling terms and $n$ rotations per 
iteration, approximately $2mn$ gates per operator. These are comparable 
as long as $m$ is less than quadratic in $n$. However, unlike the quantum 
chaotic operators, the coupling terms at every iteration of the PR 
algorithm can be applied simultaneously. The PR operators thus
require only $2m+1$ gates which appears to be less then that required of chaotic 
operators.

In conclusion, we have studied the entanglement generation of interpolating 
ensemble, PR, and quantum chaotic operators as a function
of time. These operators restrict the full space of CUE in different ways and 
the effect of these restrictions can be seen in the entanglement generation. 
The statistical properties which influence the entanglement generation include 
the number variance, which effects the entanglement generation as a function 
of time, and the matrix element distribution, which determines other aspects 
of the entanglement. Finally, we note that the PR operators may be 
efficiently implemented on a quantum computer and provide a way to create 
highly entangled, random states.

The authors thank K. Zyczkowski for clarifying interpolating ensemble
generation, J. Emerson for helpful discussions and acknowledge 
support from the DARPA QuIST (MIPR 02 N699-00) program. YSW acknowledges 
support of the National Research Council through the Naval Research Laboratory.
Computations were performed at the ASC DoD Major Shared Resource Center.


\begin{thebibliography}{99}
\bibitem{corr}
Due to misprints contained in the referenced papers \cite{Zyc2,Zyc3} 
we explicitly present the algorithm actually used in those works.
\end{thebibliography}
\end{document}